\documentclass[doublecol]{epl2} 
\usepackage{graphicx}
\usepackage{dcolumn}
\usepackage{bm}

\newcommand{\Ab}{A^{\epsilon}_L}

\newcommand{\eps}{\varepsilon}

\newcommand{\BEQ}{\begin{equation}}
\newcommand{\EEQ}{\end{equation}}
\newcommand{\BEA}{\begin{eqnarray}}
\newcommand{\EEA}{\end{eqnarray}}

\newcommand{\nn}{\nonumber}
\title{Linear Response and Fluctuation Dissipation Theorem  for\\
non-Poissonian Renewal Processes.}
\shorttitle{Title} 
\author{G. Aquino\inst{1 \footnote{E-mail: gaquino@mpipks-dresden.mpg.de}}, P. Grigolini \inst{2,3} and  B. J. West \inst{4}}

\institute{                    
  \inst{1}Max-Planck-Institut f\"ur Physik komplexer Systeme-
N\"othnitzer Str. 38
01187 Dresden, Germany \\
  \inst{2}Center for Nonlinear Science, University of North Texas,
Denton, Texas\\
  \inst{3}Dipartimento di Fisica ``E. Fermi", Universita' di Pisa-
Largo Pontecorvo 56127, Pisa, Italy \\
\inst{4} Phyics  Department, Duke University, Durham, 27708 USA
}

\abstract{
The Continuous Time Random Walk (CTRW) formalism is used to model the
non-Poisson relaxation of a system response to  perturbation.
Two  mechanisms to perturb the system are analyzed: a first in which
 the perturbation, seen as a potential gradient, simply introduces a bias in the
hopping probability of the walker from on site to the other but leaves unchanged
the occurrence times of the attempted  jumps (``events") 
and  a second in which
the  occurrence times of the events are perturbed.
 The system response is calculated analytically in both cases in  a non-ergodic
condition, i.e. for a diverging first moment in time. Two
different Fluctuation-Dissipation Theorems (FDTs), one for each kind of
mechanism,  are derived and discussed.}
\begin{document}

\maketitle

\section{Introduction}
Complex physical systems provide a number of challenges that cannot be
addressed with  the traditional approaches of equilibrium statistical
physics. One phenomenon that reveals a striking departure from traditional
statistical mechanics is Blinking Quantum Dots (BQD) \cite{bqd}. BQD systems
jump between two states, "light on" and "light off", thereby being a
realization of dichotomous fluctuations and this two-state model of BQD proves
to be sufficient to satisfactorily  explain the observed data.
There is 
consensus that these
fluctuations are simultaneously non-Poissonian and renewal
\cite{renewal1,renewal2}. This property makes them a paradigmatic example of
weak ergodicity breakdown \cite{weakergodicitybreakdown}. The Kubo-Anderson
spectral diffusion theory \cite{kubanderson} is violated \cite{barkaisilbey},
and a new theoretical approach has to be created to study the emission and
absorption processes \cite{gerardoluigi}.

Note that the theoretical approach behind the work of
Refs. \cite{weakergodicitybreakdown,barkaisilbey,gerardoluigi} is inspired to
the Continuous Time Random Walk (CTRW) theory \cite{ctrw}.
 However, it is not yet clear if
alternative approaches based on the adoption of a Hamiltonian prescription
are possible. 
Under the simplifying
assumption that the ``light on" and the ``light off" states have the same
waiting time distribution $\psi(t)$
the theoretical CTRW  prescriptions are equivalent to the following Generalized Master Equation (GME):
\begin{equation} \label{gle}
\frac{d}{dt} {\bf   p}(t) = -\int_{0}^{t}  \Phi(t-t') {\bf   K} {\bf   .}{\bf   p}(t') dt',
\end{equation}
where ${\bf p} (t)$ is a  two-dimensional vector, whose components $p_+(t)$ and
$p_-(t)$ denote the probability for the system to be in the ``on" and
``off" state, respectively. 
 The matrix ${\bf K}$ is defined by 
\BEQ\label{firsttime}
{\it {\bf K}}=\left(
\begin{array}{cc}
1 & -1\\
-1 & 1\\
\end{array}\right)
\EEQ
and the memory kernel function $\Phi(t)$ is connected to $\psi(t)$ by means of
their Laplace Transform as:
$\hat{\Phi}(u) = u\;  \hat{\psi}(u)/(1 - \hat{\psi}(u)).$
The formal equivalence between CTRW and GME was originally established by
Kenkre et al.
\cite{kms} and extended to include renewal
aging by Allegrini, Aquino
 et al. \cite{thesis, thesis2}.

Is the GME-CTRW equivalence substantial as well as merely formal? This is a question
of great importance for statistical physics. It is well known \cite{kenkrino} that
starting from a Hamiltonian description of a system plus environment, that a
Liouville equation formalism can be used to project the dynamics onto that of
the system alone. This projected representation is equivalent to the GME.

We are convinced that the question of the physical equivalence between GME and
CTRW is still not yet settled, and  in this letter, by deriving an exact
linear response result, using only CTRW arguments,  a step towards solving
this controversy is taken. We notice in fact that the Fluctuation-Dissipation
Theorem (FDT) of the  first kind \cite{kubo}, which has a Hamiltonian
origin, 
can be generalized \cite{crisanti}, within an approach compatible with a Hamiltonian origin, to
a non-stationary condition   with the following expression:
\begin{equation}
\label{leticia1}
R_{AB}(t,t^{\prime}) = \beta \frac{d}{dt^{\prime}} \langle A(t)
B(t^{\prime})\rangle \hspace{0.5 cm}  t>t',
\end{equation}
connecting the response function $R_{AB}(t,t^{\prime})$ to the correlation
function   $\langle A(t) B(t^{\prime})\rangle$.

In Eq.(\ref{leticia1}) $A$ denotes
the variable of interest,
$B$ is the  variable of the system
through which the external perturbation affects its dynamics
and $\beta$ the inverse temperature. 

 To adapt this theoretical prediction to the BQD process, we assume $A = B =
\xi(t)$, with $\xi = +1$ and $\xi = -1$ in the state $|+\rangle$ and
$|-\rangle$, respectively and $\beta=1$. Consequently, the FDT attains the  form
\begin{equation}
\label{wrong}
R(t,t^{\prime}) = \frac{d}{dt^{\prime}} C(t,t^{\prime}),
\end{equation}
where
\begin{equation}
\label{fdt1}
C(t,t^{\prime}) = \langle\xi(t) \xi(t^{\prime})\rangle.
\end{equation}
It is important to remark that the response of the two-state system,
compatible with the CTRW approach, has been studied by Sokolov
\cite{s} as well as Sokolov and Klafter \cite{sk}, with a treatment that, as we
shall see, leads to the adoption of Eq. (\ref{wrong}). Thus, we may conclude
that the formal equivalence between CTRW and GME indeed corresponds  to 
physical equivalence. However, we notice that the theoretical predictions of
these authors all rest on the same assumption, i.e. that the perturbation
seen as a potential gradient, simply introduces a bias in the hopping
probability of the walker in moving between the sites while leaving unchanged
the occurrence times of the attempted jumps.
We name this scheme the ``phenomenological approach".


 On the other hand, the 
experimental observation of the BQD phenomenon affords information on $\psi(t)$,
which,  as we shall see, depends on two parameters, $\mu$ and $T$. The
power-law parameter $\mu$ is the complexity index and depends on the cooperative nature of the
system. The  parameter $T$ determines the time scale at which the waiting
time distribution $\psi(t)$ can be considered to be equivalent to an inverse
power law with index $\mu$. It is realistic to expect that the external perturbation has
the effect of influencing these two parameters, and especially  the parameter
$T$, given the fact that $\mu$ has a cooperative origin, and consequently is
expected to be the  more robust of the two. To calculate the system's response
to perturbation in this
scheme is a difficult problem, and the  aim of the present
article is to describe a rigorous way to solve it, within the spirit
of CTRW. 
 In this letter by calculating analytically the response to a perturbation of
$\psi(t)$ we obtain as main consequence the new FDT relation
\begin{equation}
\label{right}
R(t,t^{\prime}) = -\frac{d}{dt} C(t,t^{\prime}).
\end{equation}

\section{Response and FDT within the Phenomenological Approach}
Let us consider the process of a BQD switching between ``on" ($|+\rangle$) and
``off"  ($|-\rangle$) states.
 This process can be modeled as a random walk on  a two sites lattice. At random times determined by the distribution
$\psi(t)$ the walker attempts a jump to the other site, we call these 
attempts ``events". Whenever an event  occurs the walker can either jump to the other site 
or remain in the same site with equal  probability  $1/2$.
Within the ``phenomenological approach" the perturbation $f(t)$ simply
introduces a bias in these probabilities
 while leaving
the statistics of the events, i.e. $\psi(t)$,  unchanged. 


Thus,  under the influence of the perturbation $f(t)$  the walker 
jumps to the other state or remains in the same state with  probabilities
respectively:
\begin{eqnarray}\label{jp}
w_{\pm ,\mp}(t)=Pr(\pm \to \mp)&=&\frac{1 \pm \epsilon f(t)}{2}\\
\nonumber w_{\pm , \pm}(t)=Pr(\pm \to \pm)&=&\frac{1 \mp \epsilon f(t)}{2}
\end{eqnarray}
As  mentioned above, this case has already been solved in a GME approach in Ref.\cite{s}
and within a different approach in Ref. \cite{paolo1}.
To the end of illustrating our method  and showing its general validity   we
rederive the response in this case with a CTRW approach.
Consider a random walk on a two-site lattice which starts at time
$t=0$ with equal occupation of the two sites. 
The 
occupation vector is denoted by  $ {\bf p} $ and 
 the distribution of waiting times
for the walker between the events by $\psi(t)$. We make the assumption that:
\BEQ\label{powlaw}
\psi(\tau)=\frac{\mu-1}{T}(1+\tau/T)^{-\mu},
\EEQ
having in mind systems such as the BQDs 
whose blinking behavior
is characterized by a   residence times distribution which is a power law  with diverging first moment
(i.e. $1<\mu<2$).
The occupation at time $t$ will be determined by all the possible ways the
walker can move in the interval of time $[0,t]$  between the sites. In the
presence of the perturbation, turned on at time $t=0$,  a bias will occur.
Let us divide all the trajectories in four groups that we indicate with $g_{ij}$,
determined by the starting site $i$ at time $t=0$ and the arrival site $j$ at
time $t$, with $i,j=\pm$.
The fraction  of trajectories $g_{ij}$  starting in site $i$ at time $t=0$
and  switching to  site $j$ at time $t$ is  given by $A_{ij}(t)p_i(0)$, with:
\begin{eqnarray}\label{pp}
 A_{ij}(t)&=&  \sum_{n=0}^{\infty} \; \sum_{k_1,..k_{n}=\pm}\int_{0}^{t}dt_1\psi(t_1)w_{i,k_1}(t_1)
\cdots\\
\nn&\cdot&\int_{t_{n-1}}^t dt_n  \psi(t_n-t_{n-1}) w_{k_n,j}(t_n)\Psi(t-t_n).
\end{eqnarray}

The trajectories are then sequences of ``laminar phases" between events
in which the walker remains in the same site.
The  $\psi(t_j-t_{j-1}) dt_j$ are therefore   the probabilities of
occurrence  of an event at time $t=t_j$ without implying necessarily a change of site,
while  $\Psi(t)$ is just the survival probability, i.e. the probability of
occurrence of no event in an interval of time of length  $t$.

Let us focus on the  group $g_{++}$:
the trajectories belonging to this group,  at time $t$, are again in the
 state $|+\rangle$, so they contribute only to the population of the first
site at time $t$.
Eqs. (\ref{jp}) show that,  the first-order
contribution to the response is obtained by keeping in the terms of the
sum of Eq. (\ref{pp})  only one perturbed weight $w_{k_i,k_{i+1}}(t_{i+1})$
and leaving the unperturbed weight $1/2$ in the remaining factors . 
 This implies that  for the trajectories in the group $g_{++}$ 
characterized by the same set of $n$ events,  only the terms with the
perturbed jump probabilities
 before the last laminar phase may give an overall nonzero
contribution. In fact, consider trajectories with the same set of events
(in the group $g_{++}$) , 
for every trajectory  with a
 laminar phase weighted  with the 
perturbed jumping  probability of Eq.(\ref{jp}), there will be one 
trajectory identical in every respect   apart from  the following
laminar phase being of the same length but of opposite sign
(i.e. the walker is in the other site).
These two trajectories according to Eq. (\ref{jp}), are weighted with a jumping probability with
opposite sign   and therefore  their total contribution to  first order in
$\eps$ is null.
The only surviving contribution to  first order in $\eps$ will be the one
 with perturbed jumping
probability before the last laminar phase, the latter being  fixed by the
group considered.
The argument extends  to a generic group $ g_{ij}$,  the total contribution of
this group to the
population of site $j$ at time $t$
is therefore given by $A_{ij}(t)p_i(0)$ with
$A_{ij}(t)=A^{0}_{ij}(t)+A^{\eps}_{ij}(t)$, that is an unperturbed plus a
perturbed term.  For the response we are interested only in the contribution of
the perturbed part
which at the first order, as explained above,  simplifies to $- j A^{\eps}_L(t) p_i(0)$ with
\BEA\label{Al}
\Ab(t)&=&\sum_{n=1}^{\infty} \int_{0}^t dt_1\psi(t_1)\cdots\\
\nn&\cdot& \int_{t_{n-1}}^t dt_n \psi(t_n-t_{n-1}) \frac{\eps}{2} f(t_n)\Psi(t-t_n)
\EEA
i.e.  the contribution obtained by keeping a perturbed jumping probability
only before the last laminar phase.
Summarizing, the contribution of all the four groups of trajectories
to the population vector at time $t$ is:
\BEA\label{gg}
g_{++} &\to& -\Ab(t)  \left( \begin{array}{c}
p_+(0) \\
0   \end{array} \right) \\
\nn
g_{+-} &\to&  \;\;\Ab(t)  \left( \begin{array}{c}
0\\
p_+(0)    \end{array} \right) \\
\nn
g_{-+} &\to& -\Ab(t)  \left( \begin{array}{c}
p_-(0) \\ 
0  \end{array} \right) \\
\nn
g_{--} &\to& \;\; \Ab(t)  \left( \begin{array}{c}
0\\
p_-(0)    \end{array} \right) 
\EEA

The response of the system is obtained by calculating the difference $\Sigma$
of population between the two sites. From  (\ref{gg}) one derives:
\BEQ\label{ssigma}
\Sigma(t)=p_-(t)-p_+(t)= 2 \Ab(t).
\EEQ
Using Eq. (\ref{Al}) it follows:
\BEQ \label{sigmar}
\Sigma(t)=\eps \sum_{n=1}^{\infty}Re[\int_0^t dt_n f(t_n)  \psi_n(t_n)\Psi(t-t_n)],
\EEQ

where $\psi_n(t_n)$ is just the n-times convolution of $\psi(t)$ and the sum
runs from $n=1$ since the $n=0$ term, having no events,  does not contain any
term linear in $\eps$.
 From Eq. (\ref{sigmar})
the response function turns out to be: 
\BEQ
R(t,t')=\sum_{n=1}^{\infty} \psi_n(t')\Psi(t-t')=P(t')\Psi(t-t').
\end{equation}
The result of \cite{paolo1} and \cite{s} is then recovered.
The autocorrelation function for the unperturbed case is known to be equal to the
function $\Psi(t,t')$, see \cite{thesis2}, that is:
\BEQ\label{correl}
C(t,t')=\Psi(t,t')=\int_{t}^{\infty} dx \psi(x,t'),
\EEQ
where $\Psi(t,t')$ is  the probability that, for fixed $t'$ no events occurs
until time $t>t'$, and 
\BEQ\label{psitt}
\psi(t,t')=\psi(t)+\int_0^{t'} dx \psi(t-x)P(x)
\EEQ
is the probability distribution that, for fixed $t'$, the first next event occurs
at time $t$.
It follows that
\BEQ \label{dc}
\frac{d}{dt'}C(t,t')=\int_{t}^{\infty} dx \frac{d}{dt'}\psi(x,t').
\EEQ
From Eq. (\ref{psitt})
 \BEQ
\frac{d \psi(t,t')}{d t'}= \psi(t-t')P(t')
\EEQ
and then inserting this into Eq. (\ref{dc})
one obtains:
\BEQ
\frac{d}{dt'}C(t,t')=R(t,t'),
\EEQ
that is the general FDT derivable for a system perturbed from a canonical
equilibrium.
\section{Response and FDT in the dynamic approach}
Let us now consider a scheme in which  the bias is due 
to a perturbation of the statistics of occurrence times of  events.
We name this approach ``dynamic" because we have in mind
a dynamic model  generating the waiting time distribution
of Eq. (\ref{powlaw}) through the equation of motion
\BEQ
\dot{y}=\alpha_0 \; y^{\frac{\mu}{\mu-1}},
\EEQ
with $\alpha_0=(\mu-1)/T$, describing a particle moving on the interval $I=(0,1]$
that, everytime it gets to 1, is re-injected back inside the interval $I$ with
a random initial condition.
The arrival of the particle at the border  generates the ``events" 
that lead the random walk on the two-sites lattice.
As explained in Ref. \cite{paolo1} the perturbation in this scheme  changes the
parameter $T$ and therefore  $\alpha_0$ in the following way:
\BEQ \label{alfaper}
\alpha_{\pm}(t)=\alpha_0(1 \pm \eps f(t)),
\EEQ
where the sign $\pm$ depends on the site at which the walker is residing.
We remark  that in this scheme, when an event occurs,  the walker
jumps to the other site or
remains in the same site  with unchanged  probability equal to $1/2$.
The perturbation therefore modifies the unperturbed statistics of the events,
i.e. $\psi(t)$ as given by Eq. (\ref{powlaw}), and  introduces an ``age
dependence" in the statistics of the events breaking the time translation invariance
of the renewal process.
At  first order in the perturbation (see Ref. \cite{paolo1} ), $\psi(t)$ is
changed 
into the  following conditional probability density:
\BEA\label{fundamental0}
\psi^{\pm}_{\eps}(\tau | t')&=&\psi(\tau)[1\pm \eps f(t'+\tau)]\\
&\mp& \eps
\nn \frac{\mu-1}{T }\psi_{\mu+1}(\tau) \int_{t'}^{t'+\tau} dx f(x),
\EEA
i.e. $\psi^{\pm}_{\eps}(\tau | t')d\tau $ is the probability that, after an
event has occurred at time $t'$,
 the following  event occurs  at $t=t'+\tau$.
 $\psi_{\mu+1}(t)$ is the same as Eq. (\ref{powlaw}) but  with index
$\mu+1$.
We consider the case of harmonic perturbation, then Eq. (\ref{fundamental0})
turns into:
\BEA\label{fundamental}
\psi^{\pm}_{\eps}(\tau | t')&=&\psi(\tau)[1\pm \eps cos(\omega(t'+\tau))]\\
&\mp& \eps
\nn \frac{\mu-1}{T \omega}\psi_{\mu+1}(\tau)[sin(\omega (t'+\tau))-sin(\omega t')].
\EEA
Keeping only the first term in Eq. (\ref{fundamental}) is equivalent to an unperturbed
distribution of the events followed by a bias in the choice of the direction of
motion similar to the phenomenological approach. The second term, besides
assuring normalization breaks this equivalence entangling the perturbation
with the dynamics.
By adopting the same notation as in the phenomenological case, the contribution of
the group $g_{ij}$  to the
population of the site $j$ at time $t$, is $A_{ij}(t)p_i(0)$ with 
\begin{eqnarray}\label{Aij}
&&\nn \hspace{-1.15 cm} A_{ij}(t)=\sum_{n=0}^{\infty}\sum_{k_1,...k_{n-1}=\pm}\int_{0}^{t}dt_1\psi^{i}_{\eps}(t_1|0)\frac{1}{2}
\cdots\\
\nn &&\cdot \int_{t_{n-1}}^t dt_n 
\psi^{k_{n-1}}_{\eps}(t_n-t_{n-1}|t_{n-1})\frac{1}{2}\Psi^{j}_{\eps}(t-t_n|t_n)
\end{eqnarray}
where:
\BEQ\label{Psieps}
\Psi^{\pm}_{\eps}(t-t_n|t_n)=\int_{t}^{\infty}dx \psi^{\pm}_{\eps}(x-t_n|t_n)
\EEQ
is the conditional survival  probability of remaining in the state
$\pm$, with no events  in the  interval of
time $t-t_n$, after an event occurred at time $t_n$.
Let us rewrite the survival probability in
the following way 
\BEQ
\nn\Psi^{\pm}_{\eps}(t-t_n|t_n)= \Psi(t-t_n)\pm \eps \Psi_{\eps}(t-t_n|t_n)
\EEQ
with
\BEQ\label{Psidivision}
\Psi_{\eps}(t-t_n|t_n)=  \Psi^1(t-t_n|t_n)-\Psi^2(t-t_n|t_n),
\EEQ
where, as can be deduced from  Eqs. (\ref{Psieps}) and (\ref{fundamental}),
$\Psi(t-t_n)$ is just the unperturbed survival probability 
while $\Psi^1(t-t_n|t_n)$ and $\Psi^2(t-t_n|t_n)$
are the conditional survival probabilities originating respectively from
the first and the second term linear in $\eps$ in Eq. (\ref{fundamental}), that
is:
\BEQ
 \Psi^1(t-t_n|t_n)=Re[\int_{t}^{\infty}dx\psi(x-t_n)e^{i\omega x}]\\
\EEQ
and
\BEQ
\nonumber \Psi^2(t-t_n|t_n)=\frac{\mu-1}{T \omega}
Im[\int_{t}^{\infty}dx\psi_{\mu+1}(x-t_n)(e^{i\omega x}-e^{i \omega t_n})]
\EEQ
Adopting  analogous considerations as those used   for the phenomenological
case, it can be inferred that the  first order
contribution to the population at time $t$ is obtained considering only one
laminar phase with the  perturbed distribution of Eq. (\ref{fundamental}) while for the others keeping the unperturbed one.
Also in this case one easily realizes that the contribution to the response
comes only from those trajectories with perturbed last laminar phase, that is
 by replacing all the $\psi^k_{\eps}(t_{k+1}-t_k|t_k)$ in Eq. (\ref{Aij}) with the unperturbed 
probability $\psi(t_{k+1}-t_k)$ and considering only the contribution linear in $\eps$
coming from $\Psi^{\pm}_{\eps}(t-t'|t')$.
In this case therefore, the trajectories with no events between 0 and $t$ will
also give a  contribution to  first order in $\eps$.
 Let us consider first  only the contribution of the
trajectories with at least one event, this contribution is the same 
for both the groups $g_{ii}$ and $g_{ij}$ and amounts to  $j \Ab(t) p_i(0)$, with:
\BEQ
 \Ab(t)=\frac{\eps}{2}\sum_{n=1}^{\infty}\int_0^t dt_n \psi_n(t_n)\Psi_{\eps}(t-t_n|t_n),
\EEQ
which applies to the probability vector as:
\BEA\label{gg2}
g_{++} &\to& \;\; \Ab(t)  \left( \begin{array}{c}
p_+(0) \\
0   \end{array} \right) \\
\nn
g_{+-} &\to& -\Ab(t)  \left( \begin{array}{c}
0\\
p_+(0)    \end{array} \right) \\
\nn
g_{-+} &\to& \;\; \Ab(t)  \left( \begin{array}{c}
p_-(0) \\ 
0  \end{array} \right) \\
\nn
g_{--} &\to& -\Ab(t)  \left( \begin{array}{c}
0\\
p_-(0)    \end{array} \right). 
\EEA
Then one has to  add the contribution of the trajectories with no events
which gives:
\BEA\label{last1}
  \eps \left( \begin{array}{c}
 \Psi_{\eps}(0|t)p_+(0) \\
-\Psi_{\eps}(0|t)p_-(0)   \end{array} \right).
\EEA
In the end calculating $p_-(t)-p_+(t)$ one obtains:
\BEA\label{correctzero}
 \Sigma(t)&=&2 \Ab(t)+\Psi_{\eps}(0|t)\\
\nn &=&\eps \int_0^{t}dt' [\delta(t')+\sum_{n=1}^{\infty}\psi_n(t')]\Psi_{\eps}(t-t'|t')
\EEA
Let us consider the contribution coming only from the first term in
Eq. (\ref{Psidivision}): the Laplace  transform of its contribution to
Eq. (\ref{correctzero})  is easily calculated:
\BEQ\label{firstcontr}
\hat{\Sigma}_1(s)=\eps Re[\frac{1}{s}\left(\frac{\hat{\psi}(-i \omega)-\hat{\psi}(s-i
\omega)}{1-\hat{\psi}(s-i \omega)}\right)],
\EEQ
which for a constant perturbation, i.e.  in the limit $\omega \to 0$, gives the
asymptotic value  of epsilon. 
The contribution to Eq. (\ref{correctzero}) of the second term in
Eq. (\ref{Psidivision}) 
 amounts to
\BEA \label{correct}
\nn \hat{\Sigma}_2(s)&=&-\eps\frac{\mu-1}{T \omega}
Im\left[\frac{1
}{s(1-\hat{\psi}(s-i\omega))}\left(
\hat{\psi}_{\mu+1}(-i\omega)\right.\right. \\
&&\left.\left. -\hat{\psi}_{\mu+1}(s-i\omega) 
  -(1-\hat{\psi}_{\mu+1}(s))\right)
\right].
\EEA
Considering that
$\hat{\psi}_{\mu+1}(s)=1+T s \hat{\psi}(s)/(1-\mu)$
and  making a convenient simplification we get:
\BEA \label{correct2}
 \nn &&\hspace{-0.65cm}\hat{\Sigma}_2(s)=\eps 
Im\left[ \frac{
(i \omega) \hat{\psi}(-i\omega) +(s-i \omega) \hat{\psi}(s-i
\omega)- s \hat{\psi}(s)}{-s \omega (1-\hat{\psi}(s-i\omega))}
\right]\\
&& \hspace{-0.65cm}=-\eps Re\left [\frac{\hat{\psi}(-i \omega)-\hat{\psi}(s-i\omega)}{s(1-\hat{\psi}(s-i\omega))} +\frac{\hat{\psi}(s)-\hat{\psi}(s-i\omega)}{i \omega(1- \hat{\psi}(s-i\omega))}\right].
\EEA
Taking  the limit for $\omega \to 0$ carefully so as to obtain the case of a constant
perturbation, and then taking the limit $s \to 0$ (i.e. $t \to \infty$), in
Eq. (\ref{correct2}),  the final contribution amounts to $\eps(-\mu)$. This contribution  added to that
of the first term  given by (\ref{firstcontr}) gives the value $\eps(1-\mu)$.
We confirm this result  with a  numerical simulation for the case
of constant perturbation, the good agreement is  shown in Fig. 1.\\
\begin{figure}
\includegraphics[width=7.8 cm, height=5 cm]{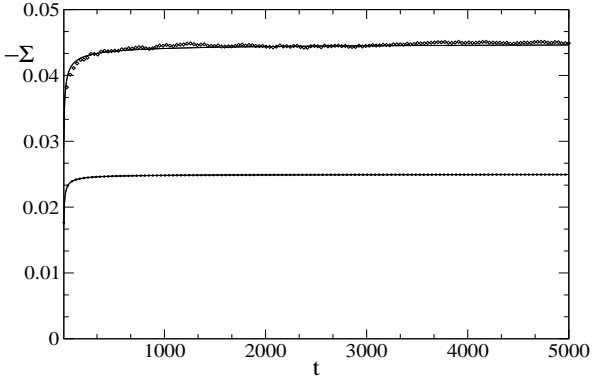}
\caption{\label{fig1}  Numerical simulation (dots) for $\Sigma(t)$ in response
to a constant perturbation  as compared to the  inverse Laplace transform of
the analytical  expression  Eq. (\ref{Lapltotal}) in the limit $\omega \to 0$. From top to bottom
the curves refer to $\psi(t)=\frac{\mu-1}{T}(1+t/T)^{-\mu}$ with  $\mu =
1.45$,  $T=1.6$ and  $\mu = 1.25$,  $T=1.6$ ($\eps=0.1$).}
\end{figure}
Remarkably,  summing  (\ref{correct2}) and (\ref{firstcontr}) and simplifying 
\BEQ\label{Lapltotal}
\hat{\Sigma}(s)=\hat{\Sigma}_1(s)+\hat{\Sigma}_2(s)=\eps Re[\frac{
( \hat{\psi}(s-i\omega) - \hat{\psi}(s))}{i \omega (1-\hat{\psi}(s-i\omega))}],
\EEQ
which is  just  the real part of the Laplace transform of 
\BEQ\label{mauro}
 \eps \int_0^t \psi(t,t')e^{i \omega t'} dt',
\EEQ
where $\psi(t,t')$ is  given by  Eq. (\ref{psitt}).
This expression
 uniquely identifies $\psi(t,t')$ as the response function
 and  is straightforwardly extended to a
generic perturbation $f(t)$, simply by using  its Fourier representation and
following the same steps used to prove Eq. (\ref{mauro}). 
\subsection{FDT in the dynamic case}
We conclude that in the dynamic case  the response 
to 
an external   perturbation $f(t)$ can be written in the form
\BEQ
\Sigma(t)=\eps \int_0^t dt' R(t,t') f(t')
\EEQ
with  $R(t,t')=\psi(t,t')$.
It can be  seen, by differentiating  Eq. (\ref{correl})
respect to $t$,  that  in this case   a different Fluctuation Dissipation
Relation (FDR) is obtained, namely:
\BEQ
\frac{d }{dt}C(t,t')=-R(t,t').
\EEQ
Of course, if a stationary condition is reached,  time translation invariance
is fulfilled and the two FDRs are equivalent.
In order to  address the discussion of how general the new FDT may 
be, we  point out that the same result has been recently 
obtained by Allegrini at al. \cite{prlal}  using a dynamical model different 
from that of Eqs. (20) and (21), fitting only the constraint of 
yielding the distribution  of Eq. (8). On the basis of  this we 
argue that Eq. (39) might be indeed  a universal result, requiring only that 
the perturbation   changes slightly the occurrence times of the 
events of the system, thereby producing a  small change of the 
parameter $T$ of Eq. (\ref{powlaw}).

Summarizing,  in  this letter
we  analyzed the problem of the response to perturbation 
of systems with slow  relaxation properties described by
non-Poissonian renewal processes  with  diverging
characteristic time.  We considered
two different mechanism for perturbing the system.
For the  first one, which we termed phenomenological, we rederived 
the results obtained in Refs. \cite{sk,paolo1} and 
a Fluctuation Dissipation Relation  equal to the one that can be demonstrated
to be valid for Hamiltonian systems perturbed out of canonical equilibrium \cite{crisanti}. For the second approach,
 which we termed dynamical and consider more realistic because it
involved a perturbation of the statistics of the "events" generating the
random walk process, we  derived an exact
analytical expression for the response, and showed that a different FDR,
Eq. (\ref{right}) is fulfilled.
\section{Concluding Remarks}
We devote these concluding remarks to 
discussing  the possible consequence of the main result of this article:
Eq. (\ref{right}). The first important result is that the equivalence between
GME and CTRW not only is questionable, but is incorrect. This is expected to
have significant  consequences on the foundation of a theory for BQD
phenomenon. Altough the  response to external perturbation of single quantum
dots within a Hamiltonian model has been addressed in several works
(e.g. \cite{azzecca}), this has not been achieved in the BQD `'regime'', which
occurs for quantum dots embedded in a disordered environment. Our result indicates that  the derivation of the BQD phenomenon from a Hamiltonian picture is not only difficult but  impossible, thereby locating the intermittent fluorescence at the level of those emergent cooperative processes that determine reductionism failure.

 We speculate here also a second possible consequence of the main result of
this article. This is connected with the concept of an effective temperature
such as was
introduced into the physics of glassy systems.  In 1997 the authors of Refs. \cite{leticia1,leticia2}, on the basis of earlier work \cite{earlierwork} (see also \cite{nextyear}) argued that it is convenient
 to replace Eq. (\ref{leticia1}) with
 \begin{equation}
 \label{effective}
 R_{AB}(t,t^{\prime}) = \beta K(t,t^{\prime}) \frac{d}{dt^{\prime}} \langle
A(t)B(t^{\prime})\rangle  \hspace{0.3 cm} t>t^{\prime}, 
 \end{equation}
 where the function $K(t,t^{\prime})$ has the role of defining an effective
temperature depending on the age of the system 
 $t^{\prime}$.
 The effective temperature has then the role of
linking the out-of-equilibrium configurations occupied by the system during its
relaxation  at a given temperature,  to  equilibrium
configurations relative to a different temperature.
 There has been an intense research activity in this direction, and we limit ourselves to quoting some papers representative of the experimental and theoretical work done in this direction \cite{grigera,roma,jabbari,dino}, with controversial results including the report of no deviation from the FDT over several decade in frequency \cite{jabbari}.  
  Is the generalization of FDT derived in this Letter with exact dynamic
arguments compatible with the concept of an effective temperature as introduced
in  Eq. (\ref{effective})? 
 Consider Eqs. (\ref{correl}) and (\ref{psitt}) from which the following relations with
the survival probability:
\BEA
-\frac{d\Psi(t,t^\prime)}{dt}& =& \psi(t,t^{\prime})\\
\nonumber \frac{d\Psi(t,t^\prime)}{dt^{\prime}} &=& P(t^{\prime}) \Psi(t-t^{\prime}),
\EEA
are determined.
Thus, the result of this letter can be expressed in the form of Eq. (\ref{effective}) ($\beta$=$1$, $A$ = $B$ = $\xi$)
with 
\begin{equation}
K(t,t^{\prime}) = \psi(t,t^{\prime}) [P(t^{\prime}) \Psi(t-t^{\prime})]^{-1}.
\end{equation}
The introduction of an effective temperature in this case has the role
of linking  our description  to one compatible with the
ordinary FDT  obtained in Refs. \cite{s,sk}, which in turn
is directly linked to a Hamiltonian derivation.

\acknowledgments
 PG acknowledges  Welch and ARO for financial support 
through Grants no. B-1577 and no. W911NF-05-1-025.


\begin{thebibliography}{0}


\bibitem{bqd} S. A. Empedocles and M. G. Bawendi, Science {\bf 278}, 2114
(1997) R.G. Neuhauser {\it et al.}
 Phys. Rev. Lett. {\bf 85}, 3301 (2000); M. Kuno et al.  J. Chem. Phys. {\bf 112}, 3177 (2000). 

\bibitem{renewal1}   	X. Brokmann, J.-P. Hermier, G. Messin, P. Desbiolles, J.-P. Bouchaud, M. Dahan, 	 	Phys. Rev. Lett. {\bf 90} 120601 (2003). 

\bibitem{renewal2} S. Bianco, P. Grigolini, P.Paradisi, J. Chem. Phys. {\bf 123}, 174704 (2005). 


\bibitem{weakergodicitybreakdown}
G. Bel, E. Barkai, Europhys. Lett. {\bf 74}, 15 (2006); G. Bel and E. Barkai,
J. Phys. Condens. Matter {\bf 17}, S4287 (2005); G. Bel and E. Barkai,
Phys. Rev. E {\bf 73}, 016125 (2006); G. Margolin and E. Barkai {\bf 94}, 080601 (2005). 

\bibitem{kubanderson} R. Kubo, J. Phys. Soc. Jpn {\bf 9}, 935 (1954); Adv. Chem. Phys. {\bf 15}, 101 (1969); P. W. Anderson, J. Phys. Soc. Jpn. {\bf 9}, 316 (1954).

\bibitem{barkaisilbey} Y. -J. Jung, E. Barkai, and R.J. Silbey,
Chem. Phys. {\bf 284}, 181 (2002); I. Goychuk, Phys. Rev. E {\bf 70}, 016109 (2004). 

\bibitem{gerardoluigi} G. Aquino, L. Palatella, and P. Grigolini, Phys. Rev. Lett. {\bf 93}, 050601 (2004). 

\bibitem{ctrw} E. W. Montroll and G. H. Weiss, J. Math. Phys. {\bf 6}, 167 (1965). 



\bibitem{kms} V. M. Kenkre, E. W. Montroll, and M. F. Shlesinger, J. Stat. Phys. {\bf 9}, 45 (1973). 
\bibitem{thesis} P. Allegrini, G. Aquino, P. Grigolini, L. Palatella, and A. Rosa, Phys. Rev. E {\bf 68}, 056123 (2003).

\bibitem{thesis2} P. Allegrini, G. Aquino, P. Grigolini, L. Palatella,
A. Rosa and B. J. West,  Phys. Rev. E {\bf 71}, 066109 (2005).
\bibitem{kenkrino} V. M. Kenkre, in \emph{Excitation Dynamics in Molecular Crystal and Aggregates}, Springer Tracts in Modern Physics (Springer, Berlin, 1982). 




\bibitem{kubo} R. Kubo, M. Toda, N. Hashitsume, \emph{Statistical Physics II: Nonequilibrium Statistical Mechanics}, Springer-Verlag, Berlin (1985).

\bibitem{crisanti} A. Crisanti, F. Ritort, Journal of Physics A (Math. Gen.),
{\bf 36} R181-R290 (2003)


\bibitem{s} I. M. Sokolov, Phys. Rev. E {\bf 73}, 067102 (2006). 

\bibitem{sk} I. M. Sokolov and J. Klafter
Phys. Rev. Lett. {\bf 97}, 140602 (2006).



\bibitem{paolo1}P.  Allegrini, G. Ascolani, M. Bologna and P. Grigolini cond-mat/0602281

\bibitem{prlal} P. Allegrini, M. Bologna, P. Grigolini, B. J. West, Phys. Rev. Let. in press

\bibitem{azzecca}M. Gosh, R. K. Hazra and S.P. Bhattacharyya
Chem. Phys. Lett. {\bf 388}, 337 (2004) 
\bibitem{leticia1} L. F. Cugliandolo, J. Kurchan, L. Peliti,
Phys. Rev. E {\bf 55}, 3898 (1997).


\bibitem{leticia2} J. -P. Bouchaud, L. F. Cugliandolo, J. Kurchan,
M. M\'{e}zard  in "Spin Glasses and Random Fields", ed. by A. P. Young, World Scientific (Singapore 1997), cond-mat/9702070. 

\bibitem{earlierwork} L. F. Cugliandolo and J. Kurchan, Phys. Rev. Lett. {\bf 71}, 173 (1993); Phil. Mag. {\bf 71}, 501 (1995); J. Phys A: Math. Gen {\bf 27}, 5719 (1994); A. Baldassari, L. F. Cugliandolo, J. Kurchan and G. Parisi, J. Phys. A: Math. Gen. {\bf 28}, 1831 (1995). 

\bibitem{nextyear} E. Marinari, G. Parisi, F. Ricci-Tersenghi, and
J.J. Ruiz-Lorenzo, J. Phys. A {\bf 31}, L481 (1998)\


\bibitem{grigera} T. S. Grigera and N. E. Israeloff, Phys. Rev. Lett. {\bf 83}, 5038 (1999).

\bibitem{roma} F. Rom\'{a}, S. Bustingorry, P. M. Gleiser, and D. Dom\'{i}nguez, Phys. Rev. Lett. {\bf 98}, 097203 (2007). 
\bibitem{jabbari} S. Jabbari-Farouji, D. Mizuno, M. Atakhorrami, F. C. MacKintosh, C. F. Schmidt, E. Eiser, G. H. Wegdam, and D. Bonn, Phys. Rev. Lett. {\bf 98}, 108302 (2007). 
\bibitem{dino} R. Mauri and D. Leporini, Europhys. Lett. {\bf 76}, 1022 (2006). 




\end{thebibliography}
\end{document}